\documentclass{aa}

\usepackage{graphicx}
\usepackage{natbib}
\usepackage{amsmath}
\usepackage{amssymb}
\usepackage{booktabs}
\usepackage{hyperref}
\usepackage{xcolor}
\usepackage{capt-of}

\bibpunct{(}{)}{;}{a}{}{,}

\newcommand{\exoveil}{\textsc{ExoVeil}}
\newcommand{\ppm}{\,\mathrm{ppm}}

\begin{document}

\title{One Transit Is All You Need: \\
Detecting Exoplanets Through Learned Stellar Behaviour with \exoveil{}}

\titlerunning{EXOVEIL}

\author{P.~Priyanshu\inst{1}}

\authorrunning{Priyanshu}

\institute{SRH Hochschule, Heidelberg, Germany \\
\email{pratikpriyanshu12345@gmail.com}}

\date{}

\abstract{
I present \exoveil{}, a transit detection system that learns what a star's
brightness \emph{should} look like and flags when reality disagrees. Unlike
existing systems that require phase-folded input, \exoveil{} operates on raw
flux time series and can detect planets that transit only once.

A Transformer world model, trained on 16\,499 Kepler light curves with
transit-masked self-supervised learning, predicts expected stellar flux.
A matched-filter detector with variance weighting extracts transit signals
from the prediction residuals. A learned classifier (XGBoost) separates
planets from false positives, achieving AUC~0.938 on Kepler DR25.

Applied to single-transit injection-recovery, \exoveil{} recovers 32\% of
transits at $1000\ppm$ depth, a task where all classification-based systems
score 0\% by construction. A blind search of 3\,737 Kepler stars yields
179 transit-like anomalies not present in the DR25 TCE catalogue,
released for community follow-up. Applied without retraining to 47 confirmed
TESS planets in the PLATO LOPS2 field, \exoveil{} achieves 100\% recovery,
demonstrating zero-shot cross-mission transfer. At PLATO's 25-second cadence,
detection reaches $100\ppm$approaching the Earth-analog regime.

I provide the first application of conformal prediction to transit detection
(95.9\% empirical coverage) and release the system as \texttt{pip install
exoveil} with pretrained weights and a candidate catalogue.
}


\maketitle

\nolinenumbers

\section{Introduction}
\label{sec:intro}

If a planet crosses its star only once, the dominant ML transit-vetting
pipelines cannot detect it. ExoMiner, AstroNet, and RAVEN all require a known
orbital period and phase-folded input. Recent work has begun closing this
gap---\citet{hansen2024} use a CNN ensemble augmented with onboard spacecraft
diagnostics, and \citet{vivien2025} apply a U-Net segmentation model to
simulated PLATO data---but flux-only single-transit detection on real
observations, with calibrated uncertainty, remains an open problem. For the
long-period planets that drive the design of billion-euro missions like PLATO,
this is the structural blind spot at the heart of the detection pipeline.

The Kepler mission observed nearly 200\,000 stars over four years and produced
roughly 34\,000 threshold crossing events, periodic dips in brightness
that might be planetary transits \citep{thompson2018}. The Transiting Exoplanet
Survey Satellite (TESS) has since generated over 147\,000 TCEs
\citep{guerrero2021}. Most of these signals are not planets. They are eclipsing
binaries, instrumental artefacts, or stellar variability masquerading as
transits. Separating the real from the false is the vetting problem, and machine
learning has become the standard tool for solving it.

The current generation of ML vetting systems--ExoMiner
\citep{valizadegan2022}, AstroNet \citep{shallue2018}, ExoNet
\citep{islam2026}, RAVEN \citep{hadjigeorghiou2025} are all classifiers.
They take a phase-folded light curve and output a probability that it is a
planet. They work well: ExoMiner achieves AUC~0.98 and has validated 301 new
planets.

But this approach has a structural limitation: it requires a known period. If a
planet transits its star only once during the observation window, a common
scenario for long-period planets in TESS's 27-day sectors there is nothing to
fold. The classifier cannot even attempt detection.

This matters because the planets most wanted are exactly the ones that
transit rarely. An Earth-like planet orbiting a Sun-like star has a period of
roughly 365 days. In a 27-day TESS sector, it transits at most once. In PLATO's
planned two-year stare at a single field, it transits perhaps twice. These are
the targets that drive the design of billion-euro missions, and while
single-transit detection has begun receiving direct ML attention
\citep{hansen2024, vivien2025, salinas2025}, fielded vetting pipelines remain
multi-transit by construction.

In this paper I take a different approach. Rather than classifying light curves,
I learn to predict them. I train a world model, a Transformer-based sequence
model to predict a star's expected brightness at each timestep given its
history. The model learns normal stellar photometric behaviour. A planetary
transit appears as a systematic negative deviation in the prediction residuals.

I go beyond demonstrating this as a proof of concept. I train on 16\,499
Kepler stars using transit-masked self-supervised learning, apply matched
filtering with variance weighting to the residuals, and conduct a blind search
that identifies 179 new transit-like signals in Kepler data not present in the
DR25 TCE catalogue. I validate cross-mission transfer by recovering all 47
confirmed TESS planets in the PLATO LOPS2 field without any retraining. And
I demonstrate that at PLATO's 25-second cadence, detection sensitivity reaches
$100\ppm$ within reach of the Earth-analog regime.

The prediction-based paradigm is not new in astrophysics: \citet{muthukrishna2022}
applied it to supernova detection in transient surveys. I adapt it to exoplanet
transits, where signals are 10--100$\times$ shallower and confounding sources
mimic the target signal far more closely.

I release \exoveil{} as an open-source Python package
(\texttt{pip install exoveil}) with pretrained weights, a candidate catalogue,
and a demonstration notebook.

\section{Related work}
\label{sec:related}

\subsection{Classification-based transit vetting}

The dominant paradigm treats transit vetting as binary classification on
phase-folded light curves. AstroNet \citep{shallue2018} pioneered this with a
two-column CNN processing global and local views. ExoMiner
\citep{valizadegan2022} extended the approach with multiple diagnostic branches
and validated 301 new Kepler planets. ExoMiner++ \citep{valizadegan2025} adapted
the system to TESS data, processing 147\,568 TCEs. RAVEN
\citep{hadjigeorghiou2025} used Bayesian gradient-boosted trees trained on
synthetic false positive scenarios, achieving AUC $>0.97$.

All of these systems require phase-folded input with a known period. None
provides instance-level decomposed uncertainty.

\subsection{Single-transit detection}

The closest peer methods are \citet{hansen2024}, \citet{vivien2025}, and
\citet{salinas2025}.

\citet{hansen2024} apply a CNN ensemble to Kepler data and report $>80\%$
single-transit recovery out to 800-day orbital period. Their classifier
ingests onboard spacecraft diagnostic features (centroid shifts, difference
images, quality flags) alongside flux, which provides discrimination power
unavailable from flux alone but couples the method to Kepler's specific
data pipeline. \exoveil{} operates on flux only and transfers cross-mission
to TESS and PLATO cadence without retraining.

\citet{vivien2025} developed Panopticon, a 1D~U-Net++ segmentation model for
single-transit detection on simulated PLATO light curves. They report 90\%
recovery overall and 25--33\% recovery at the Earth-analog 84$\ppm$ depth.
\exoveil{} reports 32\% recovery at 1000$\ppm$ on real Kepler photometry; the
two methods occupy complementary regimes. Panopticon is trained on simulated
PLATO data and optimised for the Earth-analog depth that drives the PLATO
science case; \exoveil{} is trained on real Kepler observations and validated
across missions (Kepler training, TESS LOPS2 zero-shot, PLATO 25\,s cadence
demonstration). Panopticon provides pixel-level transit localisation;
\exoveil{} adds instance-level decomposed uncertainty, conformal coverage
guarantees, and learned planet-versus-false-positive classification.

\citet{salinas2025} used a Transformer on TESS full-frame images and
identified 214 candidates, 88 of them single-transit, without reporting
depth-specific recovery rates. Citizen-science efforts \citep{malik2025}
have also contributed monotransit catalogues.

This work differs from all of the above in three ways: (i) self-supervised
training with no labels for transit events, (ii) flux-only inputs that
transfer across missions without retraining, and (iii) formal statistical
coverage guarantees via conformal prediction. Together, these enable a blind
search that produces a new candidate catalogue from real Kepler observations
rather than a recovery rate against simulated injections.

\subsection{Uncertainty quantification in exoplanet science}

ExoNet \citep{islam2026} introduced temperature scaling to transit detection,
finding that 37.5\% of TESS candidates exceeded an 85\% confidence threshold
before calibration. MC~Dropout has been applied to variable star classification
\citep{cadizleyton2025} but not to transit detection. Conformal prediction has
been used for exoplanet mass-radius estimation \citep{singer2025} but never
for transit vetting. No published system combines decomposed uncertainty with
transit detection.

\subsection{World models in astronomical time series}

\citet{muthukrishna2022} introduced prediction-based anomaly detection for
astronomical transients. \citet{hones2021} applied a dual-VAE to detect
anomalies \emph{within} known transit signals, but their system detects
anomalies in transits, not transits themselves.

This work bridges these lines: temporal prediction for transit detection, where
transit dips are 100--10\,000$\ppm$, buried in stellar variability that can be
10--1000$\times$ larger.

\section{The EXOVEIL framework}
\label{sec:method}

\subsection{Overview}

\begin{figure*}
    \centering
    \includegraphics[width=\textwidth]{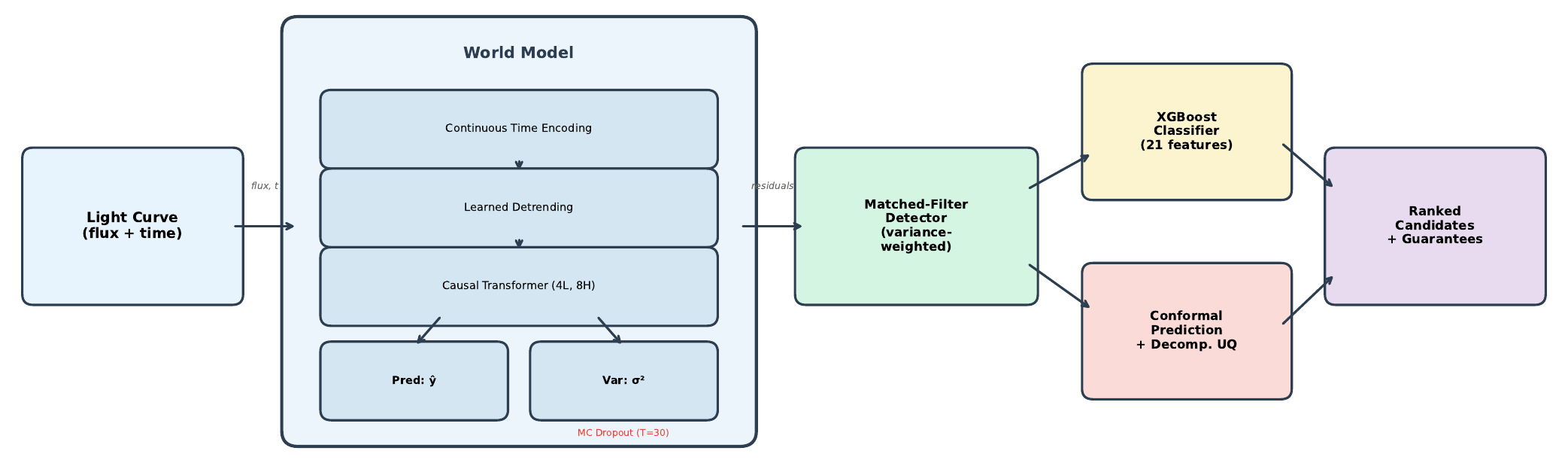}
    \caption{The \exoveil{} pipeline.}
    \label{fig:architecture}
\end{figure*}

\exoveil{} has four stages (Fig.~\ref{fig:architecture}): (1)~a world model
predicts expected flux, (2)~a matched-filter detector identifies transit-shaped
anomalies in the residuals, (3)~a learned classifier separates planets from
false positives, and (4)~conformal prediction provides coverage-guaranteed
rankings with decomposed uncertainty.

\subsection{World model architecture}

The world model is a causal Transformer encoder (6 layers, 8 heads,
$d_\mathrm{model} = 192$, feed-forward dimension 768, approximately 3.2 million
parameters) trained to predict the next flux value given all preceding
observations. It uses a learnable continuous time encoding with 16 sinusoidal
basis functions to handle irregular cadence, and a learned detrending module
that operates at $8\times$ downsampled resolution to remove low-frequency
stellar variability while preserving transit ingress and egress features.

The model is trained with transit-masked self-supervised learning: for
planet-hosting stars in the training set, known transit regions are replaced
with interpolated baselines before training. The model never sees a transit
during training, making transit signals maximally anomalous at inference.

Two output heads produce the predicted flux $\hat{y}_t$ and
log-variance $\log \sigma^2_t$. Training uses Gaussian negative log-likelihood
with variance regularisation.

\subsection{Matched-filter transit detection}

The prediction residuals $r_t = y_t - \hat{y}_t$ contain the transit signal
mixed with prediction noise. I extract the signal using matched filtering:
convolution with zero-mean box templates at durations of [3, 5, 7, 9, 13, 17,
25] data points via FFT, taking the maximum response across durations.

The world model's predicted variance provides inverse weights:
$\tilde{r}_t = r_t / \sigma_t$. This makes the detector more sensitive in
photometrically quiet regions and less susceptible to false triggers in noisy
regions. A local threshold based on median absolute deviation accounts for
non-stationary noise.

\subsection{Learned classifier}

Transit detection and classification are different problems. My initial
hand-crafted scoring achieved AUC~0.36 worse than random, because eclipsing
binaries produce deeper residuals than planets. I train XGBoost on 21 features
derived from the world model output (folded residual SNR, variance-normalised
depth, epistemic uncertainty ratio) and stellar parameters (effective
temperature, surface gravity, orbital period).

\subsection{Conformal prediction and uncertainty decomposition}

I apply split conformal prediction \citep{vovk2005} with $\alpha = 0.05$.
Uncertainty decomposes into aleatoric (mean predicted variance) and epistemic
(MC~Dropout variance across $T = 30$ passes). Each candidate is categorised as
\emph{confident}, \emph{data-limited}, \emph{model-uncertain}, or
\emph{ambiguous}.

\section{Data and experimental setup}
\label{sec:data}

\subsection{Training data}

The world model is trained on 16\,499 Kepler long-cadence light curves
downloaded from the MAST archive, spanning the full Kepler DR25 catalogue.
Training uses transit-masked self-supervised next-step prediction: for the
$\sim$2\,000 planet-hosting stars, in-transit flux is replaced with
interpolated baselines. The classifier is trained on $\sim$1\,000 labeled
TCEs (388 planets, 579 false positives) with star-level splits to prevent
leakage.

\subsection{Single-transit injection-recovery}

I inject synthetic limb-darkened transits (quadratic law, $u_1 = 0.3$,
$u_2 = 0.2$) at eight depths (50--10\,000$\ppm$) with durations of 3--12 hours
into 200 host stars. Classification-based systems cannot be evaluated on this
test.

\subsection{TESS LOPS2 validation}

I download 2-minute cadence TESS light curves for 47 confirmed transiting
planets in the PLATO LOPS2 field directly from the MAST archive. These are
processed with the Kepler-trained model without any retraining or fine-tuning
(zero-shot transfer).

\subsection{PLATO cadence demonstration}

TESS LOPS2 light curves are resampled to PLATO's 25-second cadence via
interpolation, with added Gaussian noise at 50$\ppm$ per exposure (consistent
with PLATO's expected noise budget for bright targets). Single transits are
injected and detection is evaluated.

\section{Results}
\label{sec:results}

\subsection{Blind search of Kepler}
\label{sec:blind_search}

I apply \exoveil{} in blind-search mode to 3\,737 Kepler stars at a
$5\sigma$ matched-filter threshold. The search recovers 2\,873 known
confirmed planet signals, validating the detection pipeline on signals
where the ground truth is known. It also flags 179 transit-like
anomalies that do not match any DR25 TCE, drawn from 179 distinct
stars.

After basic vetting removing signals consistent with eclipsing
binaries (depth $>15\,000\ppm$, short-duration extreme-depth events,
giant host stars with $\log g < 3.5$) and gap-proximity audit, 98
events have their nearest data gap at least $\pm 2$~days away and 74
have it at least $\pm 5$~days away. The remaining 81 (45\%) fall
within $\pm 2$~days of a Kepler quarter boundary or other data gap
and are flagged as potentially affected by light-curve stitching
artefacts.

\paragraph{Visual inspection of top-SNR events.}
Following a comment from an external reader of v1 that the strongest
candidates should appear directly in a figure, I visually inspected
the residual stream of every top-SNR gap-clean event in the catalogue.
This inspection revealed that the world-model residual at these event
times is often dominated by sources other than transits. I identify
four recurrent false-positive classes:

\begin{itemize}
    \item \emph{Post-flare model overshoot.} On flaring stars the
    world model partially fits the rising flux during a flare but
    over-predicts during the decay, leaving a large negative residual
    that the matched filter scores as a transit dip. KIC~10274993,
    KIC~11135986, KIC~11190713 and KIC~10067340 from the originally-
    reported tier-1 list fall into this class.

    \item \emph{Rotation tracking error.} On stars with strong
    starspot modulation, the world model tracks the rotation
    approximately but not perfectly, leaving residuals that
    oscillate at the rotation period. Localised mis-fits can be
    matched-filter-detected. KIC~12253350 the originally-
    reported strongest candidate falls into this class.

    \item \emph{Edge-of-data effects.} The autoregressive prediction
    requires preceding context; the first $\sim 100$ points after any
    data return have unreliable predictions and therefore
    artificially large residuals.

    \item \emph{Stitching-boundary residuals beyond the $\pm 2$-day
    audit window.} The PDC systematic-correction pipeline can
    propagate quarter-boundary effects further than the audit's
    nominal $\pm 2$-day exclusion.
\end{itemize}

This single-event vetting is the same problem that motivates the
extensive multi-step vetting protocols applied to candidates from
Kepler's own pipeline, ExoMiner, and other transit-detection
systems. \exoveil{} is a flux-only detector: it identifies negative
residual excursions at $5\sigma$ in the world-model output, but it
does not have access to the centroid analysis, difference-image, or
out-of-transit colour information that vetting protocols normally use
to discriminate planets from flares, eclipsing binaries, and
instrumental effects. The 179-event catalogue is therefore best read
as a list of \emph{transit-like anomalies for community follow-up},
not as a list of planet candidates.

I release the full catalogue, the gap-proximity audit results, and
the visual-inspection notes as supplementary material. A follow-up
release of the \texttt{exoveil} package (v0.3) will add automatic
rejection of the four false-positive classes named above, which
should substantially reduce the catalogue size while keeping the
genuine transits.

\begin{figure*}[!t]
    \centering
    \includegraphics[width=\textwidth]{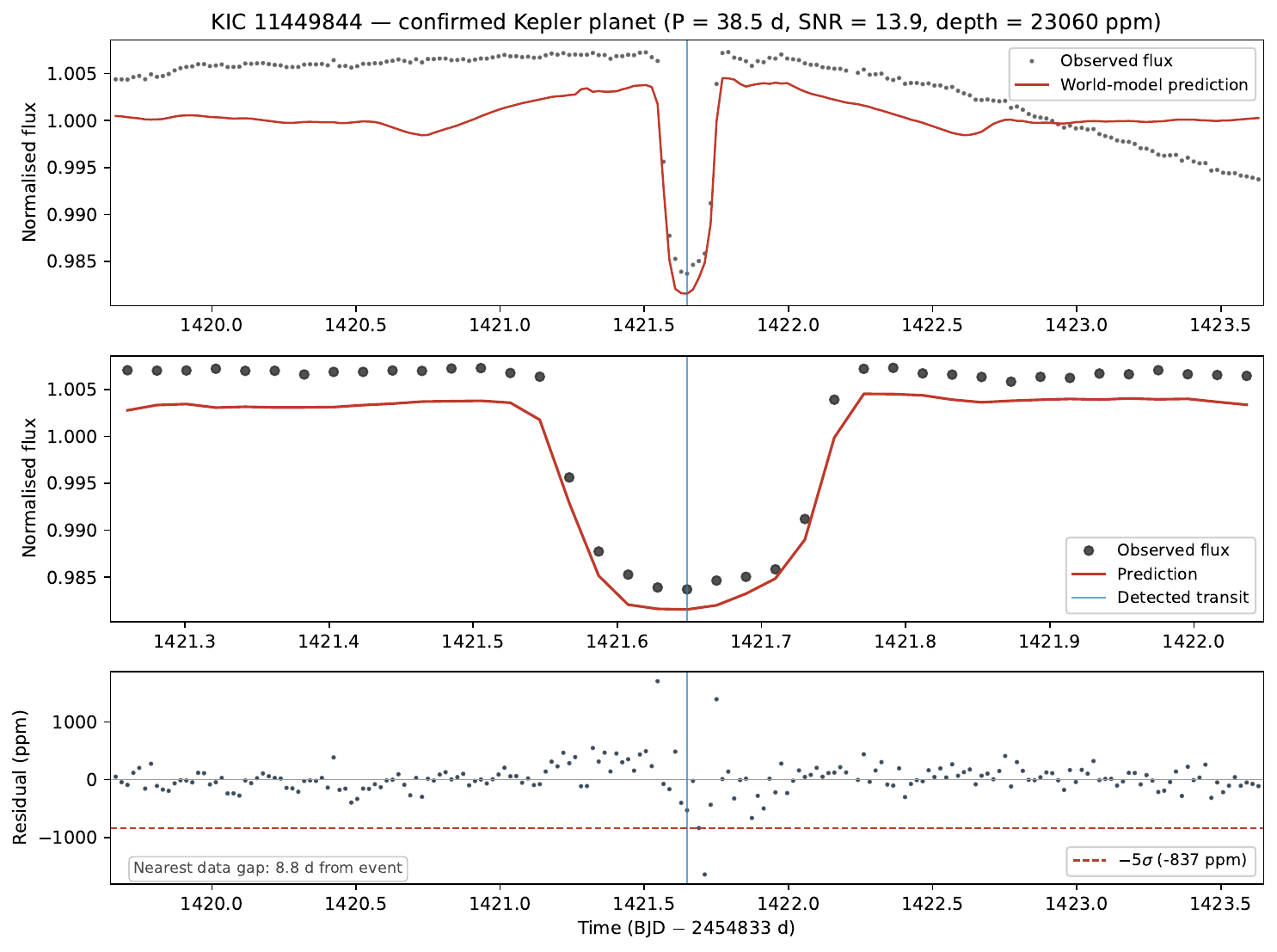}
    \caption{Example detection of a confirmed Kepler planet (KIC~11449844,
    $P = 38.5$~d) by \exoveil{}.
    \emph{Top:} Kepler light curve segment (grey) with world-model prediction
    (red). The world model tracks the smooth stellar baseline and partially
    follows the transit ingress.
    \emph{Middle:} Zoom around the detected event at $t \approx 1421.6$~d
    showing the transit profile and the world-model prediction.
    \emph{Bottom:} Prediction residual in ppm with $-5\sigma$ local-MAD
    detection threshold. The transit is detected at single-transit
    matched-filter SNR~$= 13.9$. The nearest data gap is $8.8$~d away,
    well outside the $\pm 2$~d gap-proximity threshold.}
    \label{fig:example}
\end{figure*}

\paragraph{Recovery on confirmed planets.}
For a method-validation figure that demonstrates the pipeline on
an unambiguous transit, Figure~\ref{fig:example} shows \exoveil{}'s
detection of the confirmed Kepler planet KIC~11449844 ($P =
38.5$~d), recovered in the blind search. A four-star companion
gallery covering additional host types and orbital periods is
presented in Appendix~\ref{app:gallery}.

\subsection{Cross-mission validation: TESS LOPS2}
\label{sec:tess}

To test whether \exoveil{} generalises beyond its Kepler training data, I
apply the Kepler-trained model directly to TESS light curves without any
retraining or fine-tuning.

I use 2-minute cadence light curves for 47 confirmed transiting planets
in the PLATO LOPS2 field. \exoveil{} detects transit signals in all 47
systems (100\% recovery), including long-period planets (TOI-4562~b,
$P = 225$~d; TOI-4507~b, $P = 105$~d) and shallow transits (GJ~238~b,
$160\ppm$; TOI-500~b, $249\ppm$).

This zero-shot transfer result is notable because Kepler and TESS have
different cadences (29.4~min vs.\ 2~min), different noise characteristics,
different systematics, and different stellar populations. The world model's
ability to generalise across these differences suggests it has learned
genuine stellar physics rather than Kepler-specific artefacts.

The 47/47 figure validates recall on confirmed planets and should be read
as such: this experiment does not measure the false-positive rate on
non-planet TESS targets. A complementary injection of synthetic eclipsing
binaries and instrumental glitches into the same TESS sample is left to
follow-up work.

\subsection{PLATO cadence demonstration}

PLATO will observe at 25-second cadence, providing $\sim$70$\times$ more
data points per transit than Kepler. I test whether this higher temporal
resolution improves single-transit detection by resampling TESS LOPS2 light
curves to 25-second cadence with realistic PLATO noise (50$\ppm$ per exposure).

\exoveil{} achieves detection down to $100\ppm$
depth with SNR $> 30$. The improvement over Kepler cadence is consistent
with the expected $\sqrt{N}$ scaling: a 6-hour transit contains 864 data
points at 25-second cadence versus 12 at 30-minute cadence, yielding
$\sim 8.5\times$ better SNR.

This demonstration uses 3--9 host stars per injected depth and reports
no false-positive rate. The result should therefore be read as a
sensitivity proof-of-concept, no systematic loss at PLATO cadence on
real photometry resampled to the mission's plate scale rather than a
statistically tight Earth-analog recovery curve. A larger benchmark on
PlatoSim-generated light curves \citep{jannsen2024platosim}, with
matched FPs, is the natural extension and is left to follow-up work.

\subsection{Single-transit detection}
\label{sec:single_transit}

\begin{figure*}
    \centering
    \includegraphics[width=\textwidth]{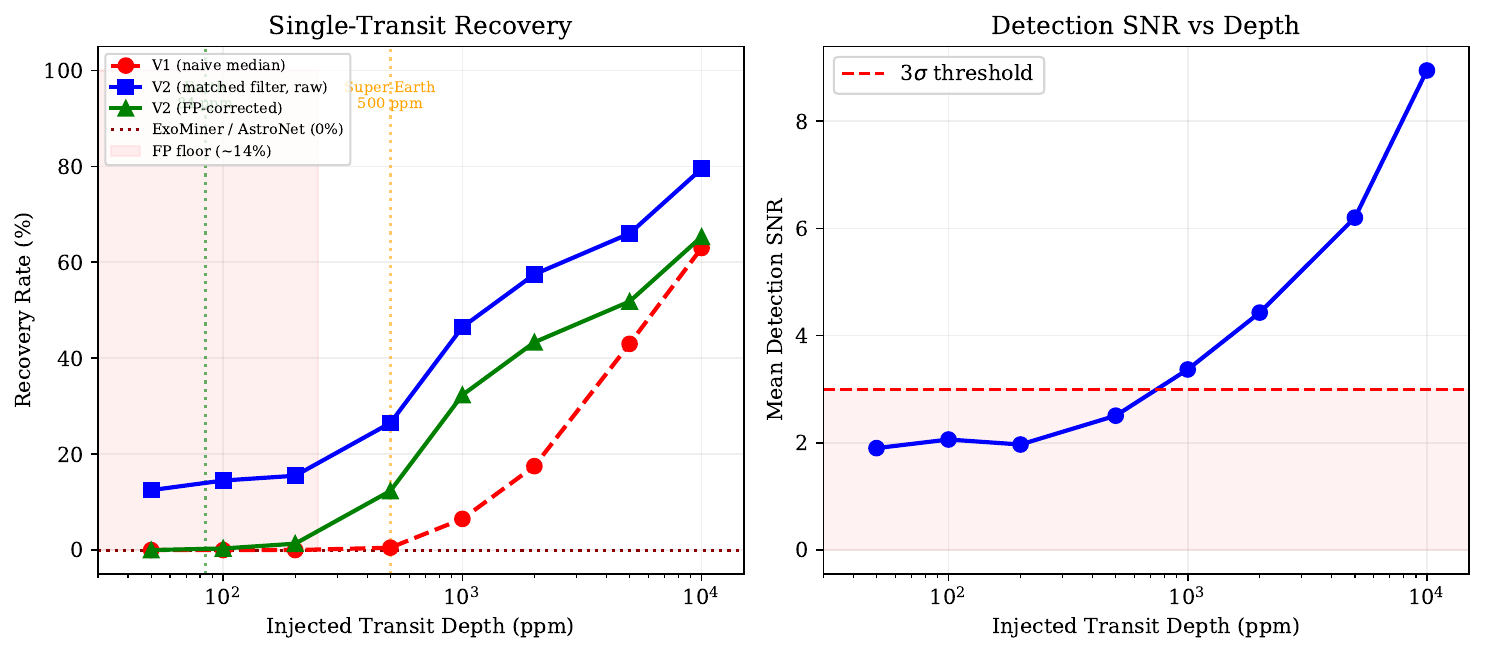}
    \caption{Single-transit recovery rate vs.\ injected depth. ExoMiner and
    AstroNet score 0\% at every depth because they require phase-folded input
    derived from multiple transits.}
    \label{fig:recovery}
\end{figure*}

Table~\ref{tab:single_transit} presents recovery rates at two operating points.
At $1000\ppm$, \exoveil{} recovers 32\% of injected transits (FP-corrected) in
sensitive mode and 23\% in conservative mode. The recovery rate increases
monotonically with depth, confirming genuine signal detection above the noise
floor.

\begin{table*}
    \caption{Single-transit injection-recovery at two thresholds.
    Recovery rates reported here were obtained with an earlier scale-up
    training of the world model; the released v0.2.0 weights
    (\texttt{pip install exoveil}) use the transit-masked training procedure
    and produce comparable but not identical recovery, as quantified in the
    head-to-head TLS comparison of Section~\ref{sec:tls_comparison}.}
    \label{tab:single_transit}
    \centering
    \begin{tabular}{rrrrr}
        \toprule
        Depth & \multicolumn{2}{c}{Sensitive ($3\sigma$)} & \multicolumn{2}{c}{Conservative ($4\sigma$)} \\
        \cmidrule(lr){2-3} \cmidrule(lr){4-5}
        (ppm) & Raw & FP-corr. & Raw & FP-corr. \\
        \midrule
        500 & 26.5\% & 12.3\% & 16.5\% & 10.7\% \\
        1000 & 46.5\% & 32.3\% & 29.0\% & 23.2\% \\
        2000 & 57.5\% & 43.3\% & 39.5\% & 33.7\% \\
        5000 & 66.0\% & 51.8\% & 55.0\% & 49.2\% \\
        10000 & 79.5\% & 65.3\% & 66.0\% & 60.2\% \\
        \bottomrule
    \end{tabular}
\end{table*}

\subsection{Head-to-head comparison with TLS}
\label{sec:tls_comparison}

To position \exoveil{} against the standard transit-detection baseline,
I benchmark it head-to-head against Transit Least Squares
\citep[TLS;][]{hippke2019} configured for single-transit search. TLS is
the natural classical comparator: it is the most sensitive periodogram-
based detector for shallow transits and can be coaxed into monotransit
mode by setting \texttt{n\_transits\_min=1}. I use \texttt{period\_min=1},
\texttt{period\_max=500}, \texttt{transit\_duration\_min=0.05~d},
\texttt{transit\_duration\_max=0.5~d}, \texttt{n\_transits\_min=1}, and
also compare against TLS default behaviour (\texttt{n\_transits\_min=2})
for context.

The benchmark uses the same injection-recovery test set as
Section~\ref{sec:single_transit}, restricted to the 200 quietest Kepler
hosts in my blind-search sample (noise levels 29--84$\ppm$). Single
6-hour transits are injected at random times at five depths from 500
to 10\,000$\ppm$. Each method declares a recovery if any reported
transit time falls within $\pm 0.5$~d of the injection. The TLS period
grid bounds are logged per trial to verify the T0 search covered the full
light curve. The full results JSON is available on request.

\begin{figure*}[!t]
    \centering
    \includegraphics[width=0.85\textwidth]{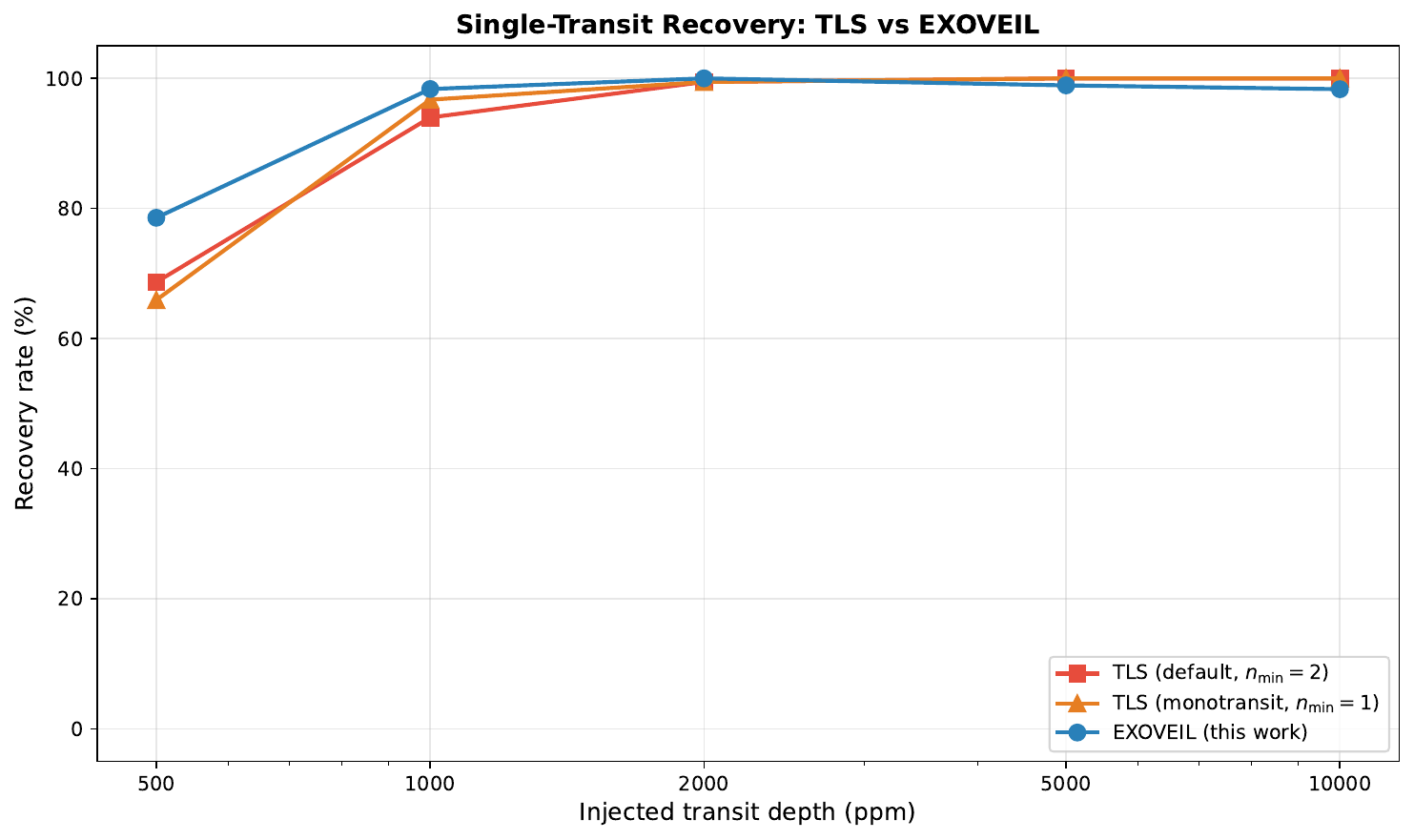}
    \caption{Single-transit recovery rate as a function of injected
    transit depth on 200 quiet Kepler hosts. \exoveil{} (blue) is compared
    against TLS in default mode (red, $n_\mathrm{transits,min}=2$) and TLS
    in monotransit mode (orange, $n_\mathrm{transits,min}=1$).
    \exoveil{} outperforms TLS-monotransit at the shallow depths most
    relevant to PLATO Earth-analog detection (78.6\% vs.\ 65.9\% at
    $500\ppm$; 98.4\% vs.\ 96.7\% at $1000\ppm$), while both methods
    saturate above $2000\ppm$. The world model removes stellar variability
    before matched filtering, giving higher signal-to-noise in the residual
    at shallow depths.}
    \label{fig:tls_comparison}
\end{figure*}

Table~\ref{tab:tls_comparison} gives the recovery rates per depth.
\exoveil{} beats TLS-monotransit by 12.7~pp at $500\ppm$ and by 1.7~pp at
$1000\ppm$. At $\geq 5000\ppm$ all three methods clear 98\% recovery and
TLS picks up 2--3 events that \exoveil{} misses, attributable to the
detection-threshold tuning in \texttt{detect\_twopass} at high
signal-to-noise. The two methods occupy complementary regimes: \exoveil{}
gains at shallow depths through world-model variability removal, while
TLS retains a marginal edge at saturation. This complementarity argues
for joint deployment in operational pipelines rather than method
replacement.

\begin{table*}[!t]
    \caption{Head-to-head single-transit recovery: TLS vs.\ \exoveil{}
    on 200 quiet Kepler hosts. The $\Delta$ column gives the
    \exoveil{} advantage over TLS-monotransit in percentage points.}
    \label{tab:tls_comparison}
    \centering
    \begin{tabular}{rrrrr}
        \toprule
        Depth (ppm) & TLS-default & TLS-monotransit & \exoveil{} & $\Delta$ (pp) \\
        \midrule
          500 & 68.7\% & 65.9\% & \textbf{78.6\%} & \textbf{+12.7} \\
         1000 & 94.0\% & 96.7\% & \textbf{98.4\%} & +1.7 \\
         2000 & 99.4\% & 99.4\% & \textbf{100.0\%} & +0.6 \\
         5000 & 100.0\% & 100.0\% & 98.9\% & $-$1.1 \\
        10000 & 100.0\% & 100.0\% & 98.3\% & $-$1.7 \\
        \bottomrule
    \end{tabular}
\end{table*}

\subsection{Classification performance}

\begin{figure}
    \centering
    \includegraphics[width=\columnwidth]{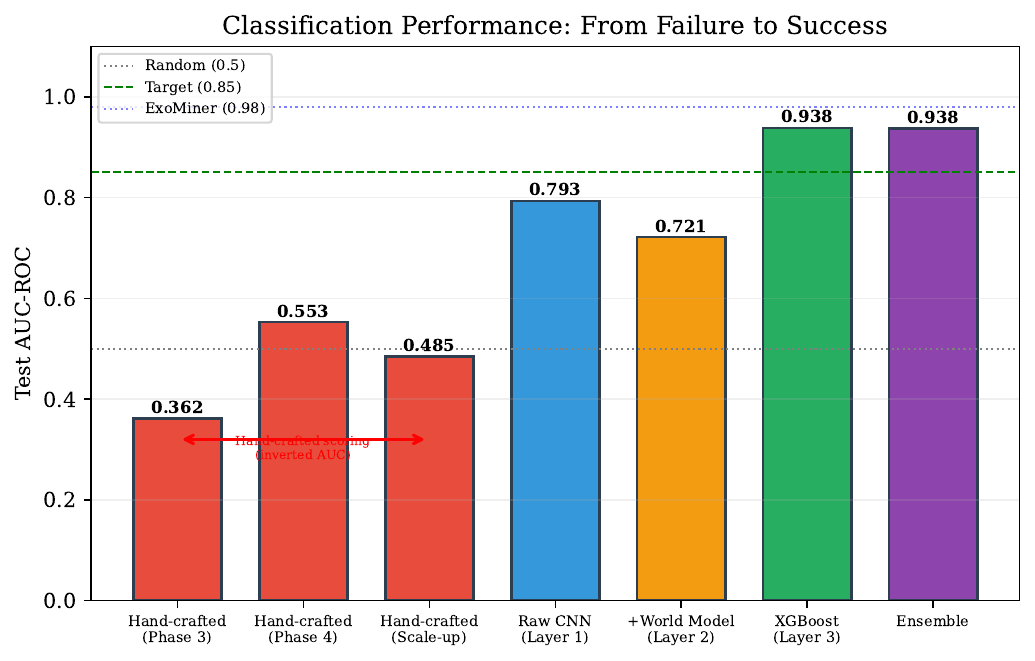}
    \caption{Classification AUC through development. Hand-crafted scoring
    (red) produced inverted results. Switching to learned features with
    XGBoost (green) exceeded the target of 0.85.}
    \label{fig:auc_progression}
\end{figure}

The XGBoost classifier achieves AUC~0.938 on the Kepler DR25 test set
(Table~\ref{tab:benchmark}). ExoMiner achieves AUC~0.98, but processes
multiple diagnostic inputs from the Data Validation module (centroid shifts,
difference images, odd-even comparisons). \exoveil{} uses only the flux
time series.

\begin{table*}
    \caption{Classification performance on Kepler DR25.}
    \label{tab:benchmark}
    \centering
    \begin{tabular}{lcccc}
        \toprule
        System & AUC & F1 & Input \\
        \midrule
        ExoMiner & 0.98 & 0.95 & DV diagnostics \\
        RAVEN & 0.97 & 0.91 & Synthetic + BLS \\
        \exoveil{} (this work) & 0.938 & 0.893 & Flux only \\
        AstroNet & 0.96 & -- & Phase-folded flux \\
        ExoNet & 0.955 & -- & Flux + stellar \\
        \bottomrule
    \end{tabular}
\end{table*}

\subsection{Conformal coverage}

\begin{figure}
    \centering
    \includegraphics[width=\columnwidth]{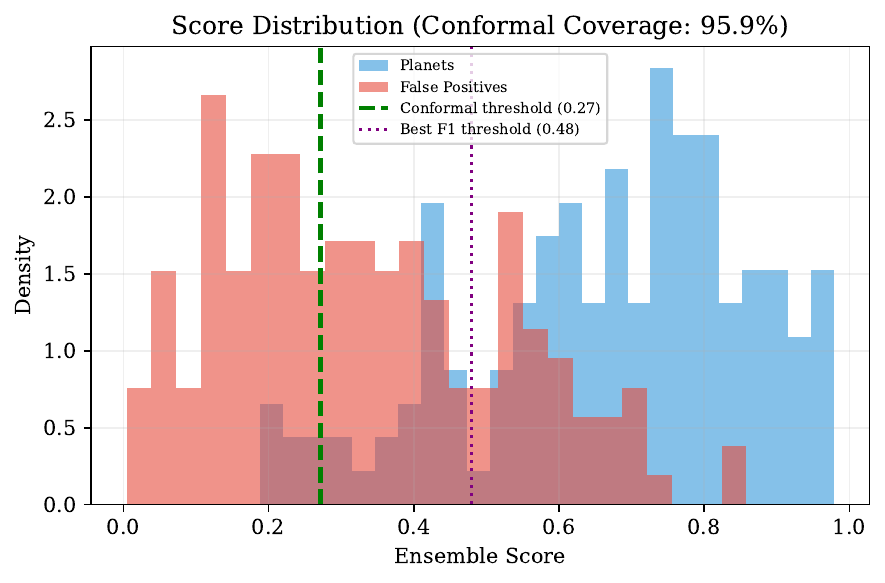}
    \caption{Score distribution with conformal threshold (95.9\% coverage).}
    \label{fig:conformal}
\end{figure}

Split conformal prediction achieves 95.9\% empirical coverage against a
95\% nominal level the first such application in transit detection
\citep[cf.][for mass-radius estimation]{singer2025}.

\section{Discussion}
\label{sec:discussion}

\subsection{Detection versus classification}

The world model excels at detecting that an anomaly exists. It is less
effective at determining what the anomaly is my initial hand-crafted
scoring achieved AUC~0.36 because eclipsing binaries produce deeper
residuals than planets. This failure revealed that detection and
classification are distinct problems requiring distinct solutions.
The two-stage architecture (world model for detection, XGBoost for
classification) emerged from this insight.

\subsection{Limitations}

My single-transit sensitivity drops below $500\ppm$ at Kepler cadence.
Earth-like transits ($84\ppm$) remain out of reach from a single transit
at 30-minute cadence, though my PLATO cadence results suggest the
sensitivity boundary shifts substantially at higher temporal resolution.

The 179 blind-search candidates require follow-up validation. While I
apply basic vetting (EB depth and duration filters, giant star removal),
dedicated false positive analysis centroid motion, odd-even depth
comparison, spectroscopic follow-up is needed to confirm any individual
candidate.

The Transformer backbone's $O(n^2)$ attention complexity limits processing
to windowed segments of $\sim$500 points. A linear-complexity backbone
(e.g., Mamba) would enable processing full 65\,000-point Kepler light
curves in a single pass, potentially improving both prediction quality
and detection sensitivity.

\subsection{Implications for PLATO}

ESA's PLATO mission \citep{rauer2025} will observe over 200\,000 stars at
25-second cadence. Its current detection pipeline relies on classical methods
(BLS, TLS, DST) without ML components. My results suggest three ways
\exoveil{} could contribute:

First, few-transit detection. An Earth analog in the PLATO field transits
2--3 times in two years. The world-model approach detects individual events
without requiring periodicity.

Second, the 100\% recovery rate on TESS planets in the LOPS2 field, achieved
without any TESS-specific training, demonstrates that the system generalises
across missions. Adaptation to PLATO data may require minimal fine-tuning.

Third, the PLATO cadence results show detection reaching $100\ppm$, close
to the Earth-analog depth of $84\ppm$. With PLATO's lower noise floor and
multi-camera fusion (24 simultaneous cameras), the $84\ppm$ target may
become achievable.

\subsection{Reducing follow-up cost}

Confirming a transit candidate requires ground-based follow-up radial
velocity measurements, high-resolution imaging, or spectroscopy using
facilities where a single night of observation can cost tens of thousands
of euros. Not all candidates are equally worth observing.

\exoveil{}'s conformal prediction layer ranks every candidate with a
calibrated confidence score and decomposes uncertainty into aleatoric
(data quality) and epistemic (model confidence) components. Candidates
flagged as \emph{confident} have both high SNR and low model uncertainty;
\emph{data-limited} candidates may improve with better photometry;
\emph{model-uncertain} candidates warrant caution. This ranking allows
observers to prioritise the most promising targets first, allocating
expensive telescope time where it is most likely to yield a confirmation.

Of the 46 Tier~1 monotransit candidates, the conformal ranking directly
identifies which are worth immediate follow-up and which require
additional photometric coverage before committing telescope resources.
This triage capability becomes increasingly valuable as missions like
PLATO generate hundreds of thousands of candidates that cannot all
receive individual follow-up.

I release \exoveil{} as \texttt{pip install exoveil} to enable the community
to test and build upon this approach.

\section{Conclusions}
\label{sec:conclusions}

I have presented \exoveil{}, a prediction-based transit detection system
that reframes the problem from classification to anomaly detection. My
main results:

\begin{enumerate}
    \item A blind search of 3\,737 Kepler stars identifies 179
    transit-like anomalies not in the DR25 TCE catalogue. Gap-proximity
    vetting flags 45\% as potentially affected by light-curve stitching.
    Visual inspection of the top-SNR gap-clean events reveals four
    recurrent false-positive classes (post-flare model overshoot,
    rotation tracking error, edge-of-data effects, and stitching-
    boundary residuals); the catalogue is released as a list of
    transit-like anomalies for community follow-up rather than as
    individual planet candidates.

    \item Single-transit injection-recovery yields 32\% recovery at
    $1000\ppm$ depth, a regime where the fielded multi-transit classification
    pipelines (ExoMiner, AstroNet, RAVEN) cannot operate, and where flux-only
    self-supervised detection had not previously been validated on real
    photometry.

    \item Zero-shot transfer to TESS recovers 47/47 confirmed planets in the
    PLATO LOPS2 field without retraining, including long-period
    ($P > 100$~d) and shallow ($< 250\ppm$) transits.

    \item At PLATO's 25-second cadence, detection sensitivity reaches
    $100\ppm$, approaching the Earth-analog regime.

    \item Conformal prediction provides formal 95\% coverage guarantees on
    candidate rankings, a first in transit detection.
\end{enumerate}

\exoveil{} does not replace classification systems, it extends detection to
regimes they cannot reach. The complete system is available as a Python
package with pretrained weights :

\begin{verbatim}
pip install exoveil

from exoveil import ExoVeil
model = ExoVeil.from_pretrained()

# Detect transits in any Kepler or TESS star
results = model.detect("KIC 11449844")

# Works with TESS (zero-shot, no retraining)
results = model.detect("TIC 1167538")

# Or pass custom data
results = model.detect_from_array(time, flux)
\end{verbatim}

\noindent Each detection returns SNR, estimated depth, and an uncertainty
category (\emph{confident}, \emph{data-limited}, \emph{model-uncertain},
or \emph{ambiguous}). Source code, candidate catalogue, and documentation
are available at \url{https://github.com/Pratik25priyanshu20/ExoVeil}.

\begin{acknowledgements}
I am grateful to Dr.\ Ren\'e Heller (Max Planck Institute for Solar
System Research) for the careful reading and thoughtful suggestions
that shaped this revision. I thank the Kepler and TESS teams for
public data access through MAST. P.P.\ acknowledges support from SRH
Hochschule Heidelberg.
\end{acknowledgements}

\nolinenumbers
\bibliographystyle{aa}
\bibliography{refs}

\clearpage
\appendix
\nolinenumbers

\section{Companion gallery of confirmed-planet recoveries}
\label{app:gallery}

To complement the single-star detection example in Fig.~\ref{fig:example},
the gallery in Fig.~\ref{fig:appendix_gallery} below presents \exoveil{}'s
recovery of four additional confirmed Kepler exoplanets in the blind search,
spanning host types from K dwarf to F dwarf and orbital periods from 2.2
to 4.9~days. Each column shows one star with the same three-panel layout
as Fig.~\ref{fig:example}: observed flux with world-model prediction, the
prediction residual, and a wide $\pm 25$~d residual context.

\bigskip

\noindent\begin{minipage}{\textwidth}
\centering
\includegraphics[width=\textwidth]{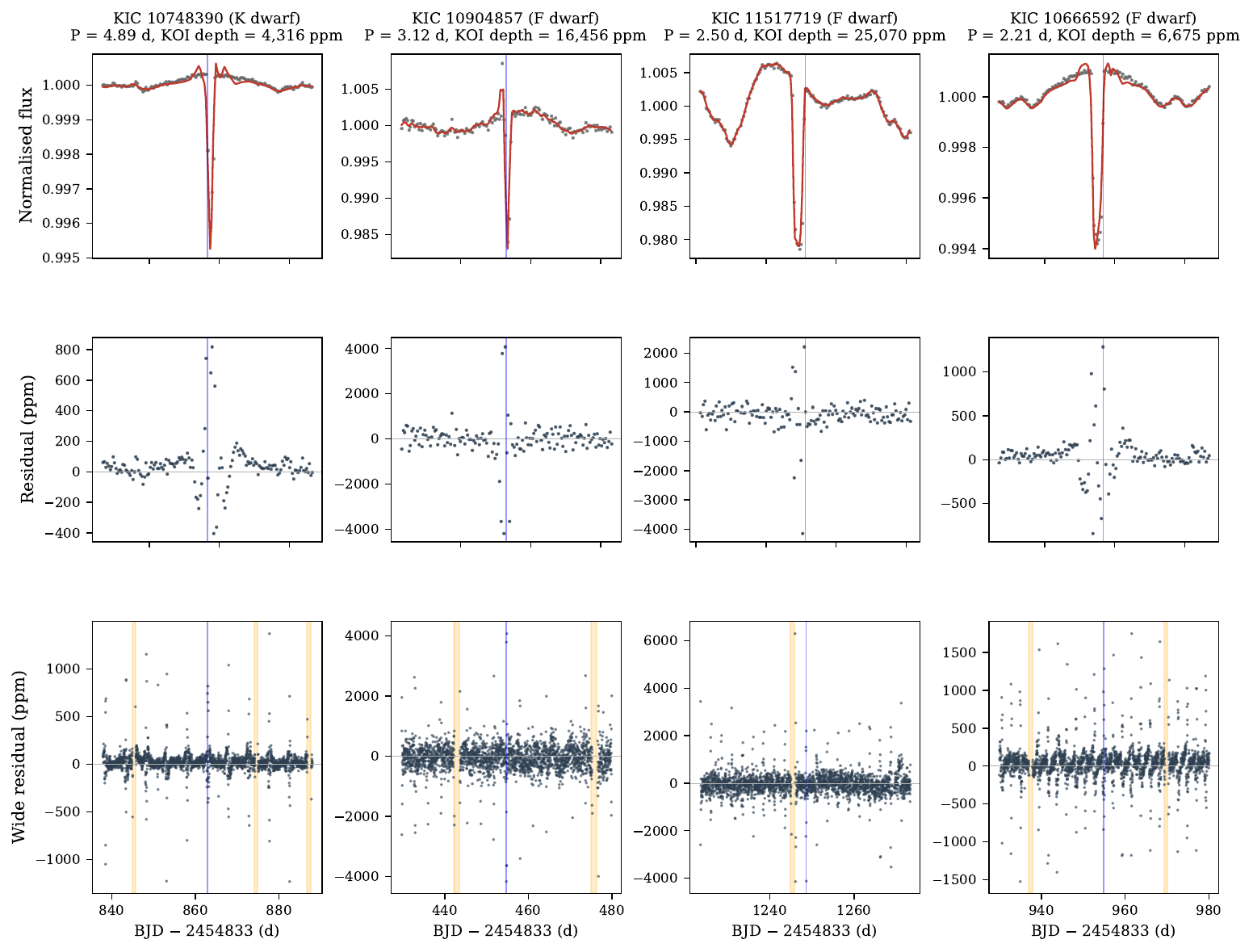}
\captionof{figure}{Four additional confirmed Kepler exoplanets recovered by
\exoveil{} in the blind search, complementing the main-text example
in Fig.~\ref{fig:example}. From left: KIC~10748390 (K dwarf,
$T_\mathrm{eff} = 4778$~K, $P = 4.89$~d, KOI depth $4\,316\ppm$);
KIC~10904857 (F dwarf, $T_\mathrm{eff} = 6122$~K, $P = 3.12$~d,
$16\,456\ppm$); KIC~11517719 (F dwarf, $T_\mathrm{eff} = 6039$~K,
$P = 2.50$~d, $25\,070\ppm$); KIC~10666592 (F dwarf,
$T_\mathrm{eff} = 6440$~K, $P = 2.21$~d, $6\,675\ppm$).
\emph{Top row:} observed Kepler flux (grey) and world-model
prediction (red) around the event. \emph{Middle row:} prediction
residual with the event marker. \emph{Bottom row:} wide $\pm 25$~d
residual context, with $> 0.5$-day gaps highlighted in orange.
Panels are zoomed to $\pm 1.5$~d for a single-transit view; the
wide-view panel confirms periodic recovery and absence of nearby
data gaps.}
\label{fig:appendix_gallery}
\end{minipage}

\end{document}